\documentclass[conference]{IEEEtran}
\usepackage{amsmath,amssymb,amsfonts}
\usepackage{algorithmic}
\usepackage{url}
\usepackage[linesnumbered,ruled,vlined]{algorithm2e}
\usepackage{graphicx}
\usepackage{textcomp}
\usepackage{xcolor}
\usepackage[colorinlistoftodos]{todonotes}
\usepackage{listings}
\usepackage{float}
\usepackage{multirow}
\usepackage{flushend}

\lstdefinelanguage{Prompt}{
    basicstyle=\ttfamily\small,
    keywordstyle=\color{blue},     % Style for keywords
    commentstyle=\color{gray},     % Style for comments
    stringstyle=\color{red!60!brown}, % Style for strings
    moredelim=[is][\color{orange}]{\{\{}{\}\}}, % Style for {{ ... }}
    breaklines=true,               % Break lines if necessary
    columns=fullflexible,
    frame=single                   % Adds a frame around the code
}

\def\BibTeX{{\rm B\kern-.05em{\sc i\kern-.025em b}\kern-.08em
    T\kern-.1667em\lower.7ex\hbox{E}\kern-.125emX}}

\newcommand{\toolname}{AugmenTest\xspace}
\newcommand{\evo}{\texttt{EvoSuite}\xspace}

\newcommand{\gptsp}{\texttt{GPT4o SP}\xspace}
\newcommand{\gptep}{\texttt{GPT4o EP}\xspace}
\newcommand{\lamasp}{\texttt{LLAMA SP}\xspace}
\newcommand{\lamaep}{\texttt{LLAMA EP}\xspace}
\newcommand{\hermessp}{\texttt{HERMES SP}\xspace}
\newcommand{\hermesep}{\texttt{HERMES EP}\xspace}
\newcommand{\raggen}{\texttt{RAG Gen}\xspace}
\newcommand{\ragsp}{\texttt{RAG SP}\xspace}
\newcommand{\ragep}{\texttt{RAG EP}\xspace}
\newcommand{\togaun}{\texttt{TOGA UN}\xspace}
\newcommand{\togawt}{\texttt{TOGA WT}\xspace}

\begin{document}

\title{\toolname: Enhancing Tests with LLM-Driven Oracles}

\author{
% \IEEEauthorblockN{Anonymous Authors}
% \IEEEauthorblockA{
% \textit{Anonymous Institution}}

\IEEEauthorblockN{Shaker Mahmud Khandaker}
\IEEEauthorblockA{\textit{Software Engineering Unit} \\
\textit{Fondazione Bruno Kessler}\\
Trento, Italy \\
skhandaker@fbk.eu}
\and
\IEEEauthorblockN{Fitsum Kifetew}
\IEEEauthorblockA{\textit{Software Engineering Unit} \\
\textit{Fondazione Bruno Kessler}\\
Trento, Italy \\
kifetew@fbk.eu}
\and
\IEEEauthorblockN{Davide Prandi}
\IEEEauthorblockA{\textit{Software Engineering Unit} \\
\textit{Fondazione Bruno Kessler}\\
Trento, Italy \\
prandi@fbk.eu}
\and
\IEEEauthorblockN{Angelo Susi}
\IEEEauthorblockA{\textit{Software Engineering Unit} \\
\textit{Fondazione Bruno Kessler}\\
Trento, Italy \\
susi@fbk.eu}
}

\maketitle

% abstract
\begin{abstract}
Automated test generation is crucial for ensuring the reliability and robustness of software applications while at the same time reducing the effort needed. While significant progress has been made in test generation research, generating valid test oracles still remains an open problem. 

To address this challenge, we present {\toolname}, an approach leveraging Large Language Models (LLMs) to infer correct test oracles based on available documentation of the software under test. Unlike most existing methods that rely on code, {\toolname} utilizes the semantic capabilities of LLMs to infer the intended behavior of a method from documentation and developer comments, without looking at the code. \toolname includes four variants: Simple Prompt, Extended Prompt, RAG with a generic prompt (without the context of class or method under test), and RAG with Simple Prompt, each offering different levels of contextual information to the LLMs. 

To evaluate our work, we selected 142 Java classes and generated multiple mutants for each. We then generated tests from these mutants, focusing only on tests that passed on the mutant but failed on the original class, to ensure that the tests effectively captured bugs. This resulted in 203 unique tests with distinct bugs, which were then used to evaluate {\toolname}. 
Results show that in the most conservative scenario, {\toolname}’s Extended Prompt consistently outperformed the Simple Prompt, achieving a success rate of 30\% for generating correct assertions. In comparison, the state-of-the-art TOGA approach achieved 8.2\%. Contrary to our expectations, the RAG-based approaches did not lead to improvements, with performance of 18.2\% success rate for the most conservative scenario. 

Our study demonstrates the potential of LLMs in improving the reliability of automated test generation tools, while also highlighting areas for future enhancement.
\end{abstract}

% keywords

% introduction
\section{Introduction} \label{sec:introduction}
In modern software development, ensuring the reliability and robustness of applications is a critical challenge, often addressed through comprehensive testing. However, automating test generation, while essential for keeping pace with rapidly evolving codebases, faces a persistent issue related to generating valid oracles \cite{earl_t__barr__2015}. This problem arises when automated tools incorrectly interpret unintended behavior as correct, leading to unreliable tests. Traditional test generation tools, such as \evo, rely on the code of the system under test to generate assertions~\cite{fraser2010mutation}. Such assertions are useful as regression oracles for identifying eventual regression bugs in successive releases by capturing deviations with respect to the previous implementation. On the other hand, however, they are not able to capture bugs due to implementations deviating from expected behavior. %, which can perpetuate these inaccuracies when bugs exist in the system.

Large Language Models, trained on vast amounts of natural language and code data, present a promising solution to this issue. By understanding both code and its associated documentation, LLMs can potentially infer the intended behavior of software, offering a more robust approach to test oracle generation. This shift from code-centric to context-aware test oracle generation could address the limitations of conventional approaches and significantly improve the usefulness of automated tests in exposing bugs.

In this paper, we introduce {\toolname}, an approach that harnesses the semantic capabilities of LLMs ~\cite{10.1145/3632971.3632973} to generate correct test oracles by leveraging code documentation and developer comments, rather than depending on the code itself. \toolname offers also the possibility to use Retrieval-Augmented Generation (RAG) which can be effective when dealing with richer contexts. {\toolname} has four distinct variants, each providing different levels of contextual information to the LLMs: Simple Prompt, Extended Prompt, RAG with a generic prompt, and RAG with Simple Prompt. 

We evaluated our approach on a dataset of 203 unique test cases for 142 Java classes, specifically selected for their ability to expose bugs. We performed experiments across multiple replications and adopted a threshold-based methodology, ensuring that consistent results were generated across different runs. We considered three thresholds, namely 60\%, 80\%, and 100\%, which reflect varying levels of result consistency across replications. These thresholds provide a measure of the reliability of the generated assertions, as explained in detail in  Section~\ref{sec:evaluation_setup}.

The results show that {\toolname}'s Extended Prompt variant significantly outperforms the other variants. Results also show that the Extended Prompt variant outperforms the state-of-the-art TOGA~\cite{10.1145/3510003.3510141} approach, achieving a 30\% success rate in the most conservative 100\% threshold scenario, compared to TOGA’s 8.2\%. However, contrary to our expectations, the RAG-based variants did not perform as well, highlighting the need for a refinement of their integration in oracle generation.

Overall, our study demonstrates the potential of LLMs to overcome the Oracle problem in software testing, offering substantial improvements over existing methods. However, the mixed results for RAG-based variants highlight opportunities for future work, particularly in improving how structured data is integrated into the generation process. By exploring the capabilities and limitations of LLMs in test oracle generation, this paper highlights the potential for these models to overcome key challenges in automated testing, paving the way for more reliable and efficient test generation tools.

The main contributions of this study are as follows:

\textbf{Contribution 1: Empirical Study and Benchmarking with Mutants:} We conducted a comprehensive evaluation of {\toolname} across test cases from 142 Java classes, analyzing its performance in generating accurate test oracles by leveraging large language models. This study uniquely benchmarks {\toolname}'s ability to infer correct assertions by using mutants (buggy versions of the classes), ensuring that the inferred oracles pass on the original class and fail on the buggy class. This empirical evaluation provides a robust framework for assessing the accuracy and effectiveness of LLM-driven oracle generation.

\textbf{Contribution 2: Comparison of Prompt Variants:} We explored multiple prompt variants, including Simple Prompts, Extended Prompts, and Retrieval-Augmented Generation, to investigate the impact of context-rich prompts on assertion inference. Our results demonstrate that while providing extended context can improve oracle generation, using RAG-based approaches does not enhance performance as expected.

\textbf{Contribution 3: Flexible Framework for LLM Oracle Inference:} Unlike previous studies, such as TOGA, which trains models to infer test oracles, {\toolname} presents a more flexible, model-agnostic framework. Our approach can be applied to any LLM, making it adaptable to evolving models and new developments in AI.

\textbf{Contribution 4: Dataset:} We are making our dataset publicly available~\cite{replication_package} to support future research on LLM-based test oracle generation. The dataset includes 203 test cases from 142 Java classes, each with at least 30 characters of developer comments per method, ensuring sufficient context for LLM-based inference. We also include the generated mutants (buggy versions of the classes), which were used to benchmark the performance of {\toolname} in inferring correct assertions.

The remainder of this paper is organized as follows: Section~\ref{sec:related_works} discusses closely related works in the area of LLMs for assertion generation and test generation in general. Section~\ref{sec:approach} presents details of the \toolname approach. A discussion of the experiment we performed to evaluate the effectiveness of \toolname, the results obtained, and the relevant threats to validity are presented in Section~\ref{sec:evaluation_setup}. Finally in Section~\ref{sec:conclusion} we present concluding remarks and outline potential directions for future work.

% related work
\section{Related Work} \label{sec:related_works}
We conducted a literature review focusing on the intersection of software testing and large language models (LLMs). Using a broad search query in the Scopus database \cite{Elsevier}, we initially identified 191 studies. After manually filtering out irrelevant papers and artifacts, we narrowed the set to 8 studies based on specific selection criteria. The selection criteria involved including studies that defined or experimented with software testing using large language models (LLMs), presented empirical findings, applied LLMs throughout the testing lifecycle, underwent peer review, and were written in English; conversely, papers were excluded if they did not focus on software testing tasks, lacked active use of LLMs, only mentioned LLMs in future contexts, were published before 2015, were secondary studies, or were unavailable as full text. Finally, using a snowballing approach, we reviewed references and added 3 more studies, resulting in 11 key papers relevant to our research.

Our related work is divided into two key areas: Oracle Generation and Unit Test Generation. Below is a summary of the most relevant studies within these categories.

\subsection{Oracle Generation}
Gabriel Ryan et al. introduce TOGA~\cite{10.1145/3510003.3510141}, a transformer-based framework for test oracle generation. TOGA integrates \evo and utilizes an oracle classifier and assertion ranker, significantly improving the inference of assertions and exceptional behaviors. It has demonstrated superior bug-finding accuracy compared to other tools. However, Liu et al.~\cite{10.1145/3597926.3598080} highlight limitations in TOGA’s evaluation methods, proposing TEval+ as a more realistic metric, revealing that TOGA’s precision in detecting bugs is much lower when evaluated under realistic conditions.

Tufano et al.~\cite{tufano_generating_2022} leverage pretrained transformers for assert statement generation, achieving substantial improvements over ATLAS, including an 80\% boost in top-1 accuracy. Nie et al.~\cite{10.1109/ICSE48619.2023.00178} introduce TECO, which applies code semantics for oracle generation, outperforming TOGA by 82\% in exact-match accuracy. They also highlight the impact of execution re-ranking in enhancing prediction accuracy.

Other approaches, like the Information Retrieval (IR)-based method by Yu et al.~\cite{yu_automated_2022}, show that combining IR and deep learning techniques outperforms purely deep learning-based solutions such as ATLAS in assertion generation tasks. Deep learning-based approaches, such as those analyzed by Shin et al.~\cite{10609742} provide an extensive analysis of Neural Oracle Generation (NOG) models, highlighting the lack of correlation between textual similarity metrics (e.g., BLEU, ROUGE) and test adequacy metrics (e.g., code coverage, mutation score), emphasizing the need for more effective evaluation methods in oracle generation.

\subsection{Unit Test Generation}
In the domain of unit test generation, Schäfer et al. present TESTPILOT~\cite{10329992}, a system that uses LLMs for end-to-end test generation, achieving high statement coverage and effective assertion generation. Tufano et al. propose an approach~\cite{10.1145/3524481.3527220}, a model fine-tuned on real-world developer-written test cases, outperforming GPT-3 and achieving test coverage comparable to \evo.

Tang et al.~\cite{10.1109/TSE.2024.3382365} explore ChatGPT's ability to generate unit test suites, finding that while it struggles with coverage compared to \evo, it excels in readability and usability. Yuan et al.~\cite{10.1145/3660783} take this further with CHATTESTER, a tool that improves ChatGPT-generated tests by reducing compilation errors and enhancing assertion accuracy.

Xie et al. introduce ChatUniTest~\cite{10.1145/3663529.3663801}, another ChatGPT-based tool that surpasses AthenaTest and \evo in several key test generation metrics, emphasizing the efficiency of ChatGPT-based repair mechanisms.

These studies underscore the significant advancements made in both oracle and unit test generation, showcasing the evolving role of LLMs in improving automated testing workflows. While most approaches focus on pretrained models, our work distinguishes itself by incorporating context from code documentation and developer comments, ensuring more targeted and efficient test oracle generation.

% approach
% \section{{\toolname}: Flexible Framework for LLM Oracle Inference}
\section{{\toolname}: a Framework for LLM-based Oracle Inference}  \label{sec:approach}
\begin{figure*}[tb]
\includegraphics[width=1\textwidth]{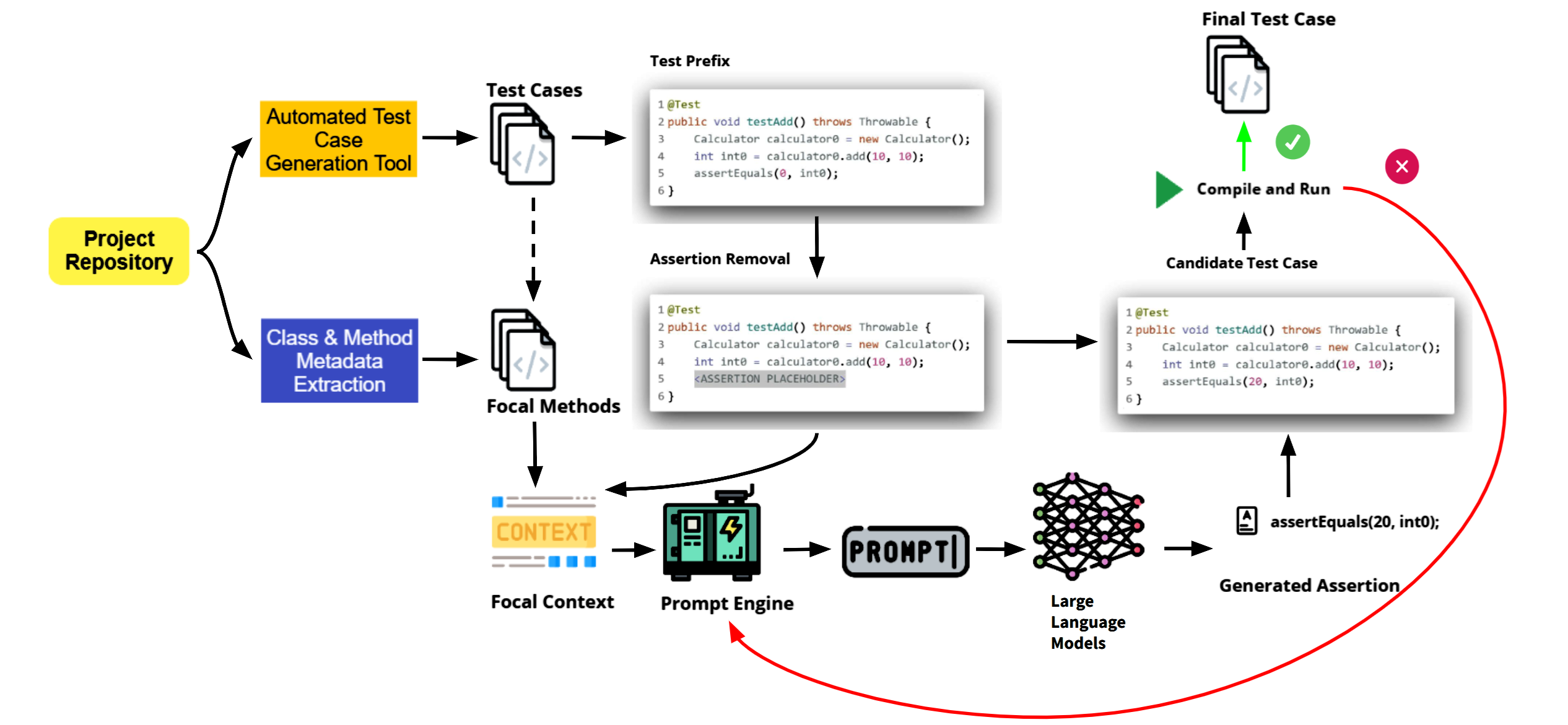}
\caption{Overview of {\toolname} approach. Each test case is processed to remove existing assertions, if any, and produce a \emph{test prefix} with a \emph{placeholder} for the eventual assertion to be generated. Then a \emph{prompt} is crafted and sent to the LLM. The returned assertion is incorporated into the test prefix, then compiled and executed . If successful the test cases is saved as a candidate, otherwise the LLM is prompted again.}
 \label{fig:Approach-overview}
\end{figure*}
In this section, we present \emph{{\toolname}}, our approach for generating test oracles using LLMs. {\toolname} tackles the problem of generating valid test oracles by leveraging the semantic understanding capabilities of LLMs to infer correct behavior based on context extracted from the Class Under Test and Method Under Test. This context includes documentation, metadata, and developer comments. \toolname also exploits  Retrieval-Augmented Generation (RAG) for structured knowledge retrieval.

\textbf{Overview of {\toolname}.} The overall workflow of {\toolname} is illustrated in Figure~\ref{fig:Approach-overview}. {\toolname} consists of two main phases: the \emph{preprocessing phase}, where relevant information from the source code is extracted and prepared, and the \emph{assertion generation phase}, where LLMs are consulted to  generate assertions based on different levels of contextual information. These two phases are presented in Algorithm~\ref{alg:oracle_generation} and described below.

% phase01
\subsection{Phase 1: Preprocessing and Metadata Extraction}
\label{subsec:phase01}
In this phase, {\toolname} analyzes the source code and extracts essential metadata about the class and method under test, building a structured knowledge base that is used later to provide context for assertion generation (lines 1-2 in Algorithm~\ref{alg:oracle_generation}). Specifically, this step includes:
\begin{itemize}
    \item Parsing each class in the project to extract metadata such as method signatures, return types, class variables, dependencies, developer comments, and relationships (e.g., inheritance or interfaces).
    \item Storing the extracted metadata in a structured or semi-structured format, which can later be used for generating contextual prompts for the LLM. For instance, in our current implementation, we store the information in \texttt{JSON} format, while in principle any other format could be adopted. 
    
    %\todo[inline, color=green]{Mentioned JSON here.}

    \item For variants using RAG (i.e., \texttt{RAG} and \texttt{RAG SP}), this information is converted into vector embeddings using a suitable embedding model. This allows efficient retrieval of relevant information during the assertion generation phase.
\end{itemize}
The result of this phase is a knowledge base containing the necessary metadata about the classes and methods in the project, along with vectorized embeddings for RAG-based retrieval when applicable.

% phase02
\subsection{Phase 2: Assertion Generation with LLMs}
\label{subsec:phase02}
In this phase, {\toolname} uses LLMs to generate test assertions based on different variants of contextual information (lines 3-24 in Algorithm~\ref{alg:oracle_generation}). This phase involves performing different tasks, as outlined below.

\textbf{Test Case Preprocessing.} For each test case, {\toolname} first identifies the \emph{focal methods} (the method which is the subject of the specific test case under consideration) in the test case and strips the assertions, leaving a \emph{test prefix}. The test prefix is prepared by keeping only the non-assertion statements in the test case and leaving a placeholder for the final assertion to be generated. Note that here we are assuming the test generation tool (e.g., \evo) generates one or more assertions together with the test cases, in which case \toolname removes them and replaces them with a placeholder which will be replaced by an assertion using the LLM. However, the preprocessing process should be equally applicable also in scenarios where the test generation tool does not generate assertions. 

\textbf{Contextual Prompt Generation.} Our prompt design prioritizes concise and deterministic outputs tailored specifically for generating JUnit assertions. Unlike traditional prompt engineering techniques such as role-playing or structured examples, which are better suited for open-ended tasks, our approach focuses on providing clear and task-relevant context (e.g., method details, developer comments) while minimizing response noise. Explicit formatting instructions (e.g., \emph{"Your statement should end with a semicolon"}) ensure compatibility with automated post-processing and downstream validation steps. This design choice optimizes efficiency and reduces ambiguity, particularly for procedural tasks, while accommodating the token and computational constraints of both closed-source APIs and local, quantized models.

Once the test case is preprocessed with a placeholder for the eventual assertion to be inserted, the next step is to formulate an appropriate \emph{prompt} to be sent to the LLM requesting for an assertion that replaces the placeholder previously inserted. The key aspects in this phase are the level of information used for constructing the \emph{context} for the prompt, which are organized into four variants as follows: 
%\todo[inline, color=green]{Added details and excerpt below.}

\begin{lstlisting}[language=Prompt, basicstyle=\ttfamily\scriptsize, breaklines=true, frame=single, float, caption={Simple Prompt Template}, label=lst:sptemp]
A Java project includes a class called {{<class_name>}} 
with following fields: {{<fields>}} and it has the 
following method details as JSON string: 
{{<focal_method_details>}}. 
We need a test oracle for a JUnit test case based on 
the above information and {{<DEVELOPER COMMENTS>}} of the 
method to test its functionality.
In the following test case, replace the 
{{<assertion_placeholder>}} with an appropriate assertion:
{{<test_method_code>}}
Just write the assertion statement for the 
placeholder, not the whole test. No explanation or 
markdown formatting/tick needed.
        \end{lstlisting}

\begin{lstlisting}[language=Prompt, basicstyle=\ttfamily\scriptsize, breaklines=true, frame=single, float,  caption={Extended Prompt Template}, label=lst:eptemp]
A Java project includes a class called {{<class_name>}} 
with following fields: {{<fields>}} and it has the 
following Focal method details as JSON string:
{{<focal_method_details>}}.
The class has other methods with developer comments as 
JSON string: {{<class_method_details>}}.
We need a test oracle for a JUnit test case based on 
the above information and {{<DEVELOPER COMMENTS>}} of the 
method to test its functionality.
In the following the test case, replace the 
{{<assertion_placeholder>}} with an appropriate assertion:
{{<test_method_code>}}
Just write the assertion statement for the 
placeholder, not the whole test. No explanation or 
markdown formatting/tick needed.
        \end{lstlisting}
        
\begin{enumerate}
    \item \textbf{Simple Prompt (\textbf{SP}):} Basic information about the class such as: \texttt{class name}, \texttt{fields} and method under test details such as: \texttt{focal method name}, \texttt{signature}, \texttt{parameters}, \texttt{dependencies}, \texttt{return type}, \texttt{developer comments} are extracted from the stored structured data and are replaced in the \emph{simple prompt template} along with the \texttt{test prefix} with \texttt{assertion placeholder} as shown in Listing~\ref{lst:sptemp}.        
    \item \textbf{Extended Prompt (\textbf{EP}):} In addition to the information added in Simple Prompt, detail about all \texttt{methods} in the class under test are added to the context to replace in the \emph{extended prompt template} as shown in Listing~\ref{lst:eptemp}.
    \item \textbf{RAG with Generic Prompt (\textbf{RAG}):} No context about class or method under test is added in the prompt. Instead, retrieval-augmented generation is used to retrieve relevant information of the class from the structured data store to generate assertions as shown in Listing~\ref{lst:raggen}.
    \item \textbf{RAG with Simple Prompt (\textbf{RAG SP}):} Combines the Simple Prompt and RAG with the database of class/method information as shown in Listing~\ref{lst:ragsimp}.        
\end{enumerate}

\begin{lstlisting}[language=Prompt, basicstyle=\ttfamily\scriptsize, breaklines=true, frame=single, float, caption={RAG Generic Prompt Template}, label=lst:raggen]
A Java project includes a class called {{<class_name>}}. 
We need a test oracle for a JUnit test case based on 
the information you can find in the provided files and 
developer comments of the method to test its 
functionality.
In the following the test case, replace the 
{{<assertion_placeholder>}} with an appropriate and 
correct assertion:
{{<test_method_code>}}
Just write the assertion statement for the 
placeholder, not the whole test. Your statement 
should end with a semicolon. No explanation or 
markdown formatting/tick needed.
        \end{lstlisting}

\begin{lstlisting}[language=Prompt, basicstyle=\ttfamily\scriptsize, breaklines=true, frame=single, float, caption={RAG Simple Prompt Template}, label=lst:ragsimp]
A Java project includes a class called {{<class_name>}} 
with following fields: {{<fields>}} and it has the 
following method details as JSON string:
{{<focal_method_details>}}.
We need a test oracle for a JUnit test case based on 
the above information and {{<DEVELOPER COMMENTS>}} of the 
method along with any other relevant information you 
can find in the provided vector store files to test 
its functionality.
In the following the test case, replace the 
{{<assertion_placeholder>}} with an appropriate assertion:
{{<test_method_code>}}
Just write the assertion statement for the 
placeholder, not the whole test. Your statement should 
end with a semicolon. No explanation or markdown 
formatting/tick needed.
        \end{lstlisting}

\textbf{Assertion Generation with LLM Interface:} Once the contextual prompt is prepared, {\toolname} uses the LLM Interface, a core component responsible for interacting with any chosen LLM to generate assertions. It is designed with flexibility and modularity in mind, enabling developers to integrate different LLMs into the test oracle generation process seamlessly. This interface abstracts the underlying model selection, ensuring that {\toolname} can work with a wide range of LLMs — whether they are open-source, closed-source, large, or quantized models. By decoupling the oracle generation process from specific LLM implementations, the interface ensures that our system remains versatile and future-proof as new LLMs are developed. 
When the selected LLM is prompted to generate an appropriate assertion for the test case, it returns a response with potential assertion based on the context provided, which is later extracted and inserted into the test prefix to complete the test case.

\textbf{Validation and Execution.} After generating an assertion and completing the test case prefix, {\toolname} checks the syntactic correctness of the test case and attempts to compile and execute it (lines 18-22 in Algorithm~\ref{alg:oracle_generation}).
% 
% \textbf{Compilation and Runtime Execution.} 
If the test case compiles successfully and executes without any errors (regardless of whether it passes or fails), the test is considered a valid candidate.
% 
% \textbf{Failure Handling.} 
If the test fails to compile or executes with errors, the assertion generation process is re-tried for a fixed number of attempts within a predefined budget until either a valid candidate is found or the budget is exhausted.

\textbf{Test Selection and Candidate Generation.} Once a valid test case is generated (i.e., it compiles and runs successfully), it is added to the pool of candidate test cases. This process is repeated for each test case in the project, with the LLM generating assertions tailored to each focal method.

In summary, the process in {\toolname} begins by extracting metadata from the project repository, preparing context for the test cases, formulating a suitable prompt for the LLM, and prompting the LLM to generate assertions based on the selected variant (SP, EP, RAG, or RAG SP). The generated assertions are validated by compiling and executing the test cases. Successful test cases are added to the candidate pool, while failed attempts are re-tried within a limited budget. 

% algorithm
\begin{algorithm}[htb]
\scriptsize % or \footnotesize or \tiny
\SetAlgoNlRelativeSize{-1} % Reduces line number size
\caption{Oracle Generation with \toolname}
\label{alg:oracle_generation}
\SetKwInOut{Input}{Input}
\SetKwInOut{Output}{Output}

\Input{Project repository/source code, Test cases (auto-generated or manual), LLM model, Variants: \texttt{SP, EP, RAG, RAG SP}}
\Output{Candidate test cases with inferred assertions}

\SetKwFunction{ExtractMetadata}{ExtractMetadata}
\SetKwFunction{ConvertToEmbeddings}{ConvertToEmbeddings}
\SetKwFunction{IdentifyFocalMethods}{IdentifyFocalMethods}
\SetKwFunction{PrepareTestPrefix}{PrepareTestPrefix}
\SetKwFunction{GenerateAssertion}{GenerateAssertion}
\SetKwFunction{CompileAndRun}{CompileAndRun}
\SetKwFunction{SelectVariantPrompt}{SelectVariantPrompt}
\SetKwFunction{ConvertToVectors}{ConvertToVectors}
\SetKwFunction{PreparePrompt}{PreparePrompt}

\BlankLine
% Phase 1: Extract Metadata and Prepare Embeddings
\raggedright 
$metadata \leftarrow \ExtractMetadata{\texttt{source code}}$\;
$embeddings \leftarrow \ConvertToVectors{metadata}$\;

\BlankLine
% Phase 2: Process Test Cases and Generate Assertions
\ForEach{$test \in test\ cases$}{
    % Step 1: Identify focal methods and prepare the test prefix
    $focalMethods \leftarrow \IdentifyFocalMethods{test}$\;
    $testPrefix \leftarrow \PrepareTestPrefix{test}$\;
    
    % Step 2: Select the appropriate prompt based on the variant
    \If{$variant == SP$}{
        $context \leftarrow$ Basic context of CUT and MUT\;
    }
    \If{$variant == EP$}{
        $context \leftarrow$ Basic context + all methods in CUT\;
    }
    \If{$variant == RAG$}{
        $context \leftarrow$ Retrieved structured JSON embeddings with no specific CUT/MUT context\;
    }
    \If{$variant == RAG SP$}{
        $context \leftarrow$ Simple prompt context + structured JSON embeddings\;
    }

    % Step 3: Generate Assertion with LLM
    $success \leftarrow False$\;
    $attempts \leftarrow 0$\;
    \While{not $success$ and $attempts < budget$}{
        $assertion \leftarrow \GenerateAssertion{context}$\;
        \If{\CompileAndRun{$testPrefix, assertion$}}{
            $success \leftarrow True$\;
            candidateTests.append($testPrefix + assertion$)\;
        }
        \Else{
            $attempts$ += $1$\;
        }
    }

    % Step 4: Store candidate test cases
    \If{$success$}{
        candidateTests.append($testPrefix + assertion$)\;
    }
}
\Return{candidateTests}\;
\normalsize % Resets font size to normal
\end{algorithm}

% evaluation
\section{Evaluation} \label{sec:evaluation_setup}
% \textbf{Research Questions.} 
To evaluate our approach, we conducted a set of experiments to answer the following research questions:

\textbf{RQ1: Can {\toolname} infer correct test oracles?}

With this question, we aim to evaluate the ability of {\toolname} to generate meaningful and accurate assertions for test cases where the original assertions, if any, are replaced with placeholders. The focus is on assessing whether the generated assertions are consistent with the expected behavior of the class and method under test, as described in the relevant documentation such as developer comments.

\textbf{RQ2: Does providing additional context improve assertion generation?}  

With this question, we aim to explore whether enhancing the context for LLM-based assertion generation by integrating  additional information, such as class information, method data, and developer comments, improves the accuracy of the inferred test oracles. Different levels of context information are provided through LLM prompt variants, and the effectiveness of each variant is measured. 

\subsection{Prototype} \label{sec:prototype}
We have implemented the \toolname approach in a prototype using \texttt{Python 3.10}, incorporating various open-source tools and libraries to enhance its modularity and flexibility. The prototype integrates two types of LLMs: a closed-source API and open-source models. For closed-source, we utilize the \texttt{OpenAI} API for \texttt{GPT-4o}\footnote{\url{https://openai.com/index/hello-gpt-4o/}}. For open-source, we employ quantized models like \texttt{Meta-Llama-3-8B-Instruct}\footnote{\url{https://huggingface.co/meta-llama/Meta-Llama-3-8B-Instruct}} and \texttt{Nous-Hermes-Llama2-13b}\footnote{\url{https://huggingface.co/NousResearch/Nous-Hermes-Llama2-13b}} using the \texttt{GPT4All}\footnote{\url{https://www.nomic.ai/gpt4all/}} (v2.7.0) framework, enabling the models to run locally and privately without relying on external servers. For our prototype, subject models do not include LLMs for code and we only considered general purpose LLMs. The choice to utilize general purpose LLMs rather than code-specific LLMs was intentional and based on the nature of the problem domain. While LLMs for code, such as \texttt{Codex}\footnote{\url{https://openai.com/index/openai-codex/}}, excels at generating syntactically correct code and understanding programming constructs due to their specialized training on code corpora, our primary focus lies in understanding and leveraging textual descriptions, such as developer comments. These descriptions provide high-level guidance on test oracle construction, which requires advanced natural language understanding — a strength of general purpose LLMs trained on broader datasets that include both code and natural language. Moreover, the generated code in our context is minimal, typically limited to expressions for assertions or oracles, where the key challenge is correctly interpreting the semantic intent conveyed in the comments. Although integrating code-specific LLMs could be explored as part of future work, our current approach prioritizes the linguistic interpretive capabilities of general purpose LLMs to align with the requirements of our methodology.

For test generation, the prototype uses \evo\footnote{\url{https://www.evosuite.org/}} (v1.2.0), an automated Java unit test generation tool. \texttt{Tree-sitter}\footnote{\url{https://tree-sitter.github.io/tree-sitter/}} (v0.20.1) is employed to parse Java code and extract metadata such as method names, classes, comments, etc. The \texttt{Jinja2} (v3.1.2) template engine is utilized to dynamically create prompts tailored for each class and method under test. \texttt{LangChain}\footnote{\url{https://www.langchain.com/ framework/}} serves as the framework to manage interactions between the code context and LLMs.

In our prototype, we employed OpenAI's \texttt{File Search Assistant}\footnote{\url{https://platform.openai.com/docs/assistants/tools/file-search}} tool to implement the RAG variants. This tool utilizes the \texttt{text-embedding-3-large} model with 256 dimensions for embedding, a chunk size of 800 tokens with a 400-token overlap, and incorporates up to 20 chunks into the context. 

Additionally, the prototype uses \texttt{JUnit} (v4.12) for compiling and executing tests, and the Major Mutation Framework\footnote{\url{https://mutation-testing.org/}} (v3.0.1) for generating mutants. The prototype was used for the experimental evaluation discussed in this section and is publicly available at: \url{https://github.com/se-fbk/augmentest}.

% Evaluation Setup
\subsection{Evaluation Setup}

\textbf{Dataset.} The evaluation is based on Java classes sampled from a dataset published in a recent study~\cite{Gruber2023} consisting of 418 Java projects with more than 27,000 classes. From this dataset, 
we selected all 142 Java classes in which every method contains developer comments of at least 30 characters (see Table~\ref{tab:dataset_summary} for the characteristics of the selected classes). A class was included only if all its methods met this criterion, ensuring that the comments provided meaningful information rather than automatically generated parameter names. To evaluate our approach, we generated multiple mutants using the Major Mutation Framework for these selected classes to introduce potential bugs. Using \evo, we generated test cases for these mutants. From these generated tests, we selected those that passed on the mutant but failed on the original class, ensuring that they accurately captured the bugs introduced by the mutation. This resulted in 203 unique test cases. Among these, 93 contained \texttt{Exception Oracles}, which ensure that expected exceptions occur or unexpected ones are avoided, and 110 contained \texttt{Assertion Oracles}~\cite{10.1145/3510003.3510141}, which verify the correctness of specific program outputs by checking conditions such as equality. The projects span various domains, including real-world examples, making the evaluation robust and potentially generalizable to actual development scenarios.

\begin{table}[tb]
\centering
\begingroup\fontsize{8pt}{9pt}\selectfont
\vspace{3mm}
\begin{tabular}{rrrrr}
  \hline
 & n. Classes & n. Loc & n. Methods & CCN \\ 
  \hline
All classes & 27,161 & 63 & 9 & 2.3 \\ 
  Selected classes & 142 & 99 & 13 & 3.0 \\ 
   \hline
\end{tabular}
\endgroup
\caption{Characteristics of the selected classes and of all the classes in Gruber et al.~\cite{Gruber2023}. \emph{n. Loc}, \emph{n. Methods}, \emph{CCN} represent average lines of code, number of methods, and cyclomatic complexity respectivey.}
\label{tab:dataset_summary}
\end{table}

% Define JSON style for listings without line numbers
\lstdefinelanguage{json}{
    basicstyle=\ttfamily\tiny,
    numbers=none, % Disable line numbers
    showstringspaces=false,
    breaklines=true,
    frame=lines,
    keywordstyle=\color{blue},
    stringstyle=\color{green!60!black},
    moredelim=[s][\bfseries\color{orange}]{:}{\ },
    morestring=[b]"
}

The RAG variants of \toolname\ rely on a structured JSON file to provide detailed metadata for Java projects. Each JSON entry corresponds to a Java class and encapsulates class-level and method-level details. Listing~\ref{lst:json_structure} shows an example.

\begin{lstlisting}[language=json, caption=Example JSON structure for RAG input, label=lst:json_structure]
{
  "projectName": "ExampleProject",
  "classes": [
    {
      "className": "ExampleClass",
      "filePath": "src/main/java/ExampleClass.java",
      "signature": "public class ExampleClass extends BaseClass implements InterfaceA",
      "superClass": "BaseClass",
      "interfaces": ["InterfaceA"],
      "package": "com.example",
      "imports": ["java.util.List", "java.io.File"],
      "methods": [
        {
          "methodName": "exampleMethod",
          "signature": "public String exampleMethod(int param1, String param2)",
          "returnType": "String",
          "visibility": "public",
          "parameters": [
            {"name": "param1", "type": "int"},
            {"name": "param2", "type": "String"}
          ],
          "comments": "This method performs example functionality."
        }
      ]
    }
  ]
}
\end{lstlisting}

The JSON file includes:
\begin{itemize}
    \item \textbf{Class-level Metadata:} Includes the class name, file path, signature, superclass, implemented interfaces, package, and imported libraries.
    \item \textbf{Method-level Metadata:} Covers method names, signatures, return types, visibility, parameters (name and type), and developer comments.
\end{itemize}

\textbf{\toolname Variants.} As stated earlier we have four different variants of \toolname each using different levels of context information in their prompts (Section~\ref{subsec:phase02}). Furthermore, different kinds of models are used for the actual assertion inference, overall resulting in the following 9 configurations of \toolname to be evaluated: 

\begin{itemize}
    \item \gptsp Simple Prompt variant using OpenAI's GPT4o model 
    \item \gptep  Extended Prompt variant using OpenAI's GPT4o model 
    \item \lamasp  Simple Prompt variant using the quantized version of Meta-Llama-3-8B-Instruct model
    \item \lamaep   Extended Prompt variant using the quantized version of Meta-Llama-3-8B-Instruct model
    \item \hermessp  Simple Prompt variant using the quantized version of Nous-Hermes-Llama2-13b model
    \item \hermesep  Extended Prompt variant using the quantized version of Nous-Hermes-Llama2-13b model
    \item \raggen   RAG with generic prompt using OpenAI's GPT4o model
    \item \ragsp  RAG with Simple Prompt using OpenAI's GPT4o model
    \item \ragep  RAG with Extended Prompt using OpenAI's GPT4o model
\end{itemize}

\textbf{Baseline.} As a baseline for comparison, we consider TOGA. TOGA ranks assertions based on a combination of its classification accuracy and confidence levels~\cite{10.1145/3510003.3510141}. It uses specific thresholds for various assertion types, such as \texttt{assertTrue} and \texttt{assertEquals}, to determine the minimum confidence required, which restricts its ability to produce useful assertions when the confidence is low. This often results in correctly classified test prefixes yielding no output~\cite{10.1145/3611643.3616265}. To mitigate this issue, we use two variants of TOGA. The first variant (\togawt) is the default TOGA tool with default configurations as publicly available. The second variant (\togaun) is a variant where we relaxed its settings by disregarding the confidence thresholds to allow the generation of assertions even at lower confidence levels. This approach enabled a more comprehensive evaluation of TOGA's capabilities and ensured that potentially valuable assertions were not excluded from consideration, thus facilitating a more balanced comparison with {\toolname}.

\toolname's performance is compared to the TOGA approach which uses custom trained neural networks~\cite{10.1145/3510003.3510141} to infer oracles. While TOGA evaluates on datasets such as Defects4J~\cite{10.1145/2610384.2628055} and Methods2Test~\cite{10.1145/3524842.3528009}, which are datasets most probably seen by most recent LLMs, potential data leakage is very likely (a concern with TOGA's reliance on previously trained datasets). By constructing a new dataset of bugs, we evaluate {\toolname}'s performance in a setting in which the system under test's code is unseen by the LLMs.

\textbf{Evaluation Metrics.} To evaluate the effectiveness of a variant in generating a correct test oracle, we compile and  execute the test case augmented with the assertion generated by the variant against the mutant (buggy) and original class from which the test case was generated. If the test passes on the original class but fails on the buggy we consider the assertion to be a correct one (a \emph{true positive}). There are other combinations of pass and fail resulting in three other outcomes. In particular, the test with the generated assertion could compile correctly but fails on both original and buggy code (\emph{false positive}); could pass on both original and buggy (\emph{true negative}); and could fail on the original but pass on the buggy (\emph{false negative}). This way of categorizing the outcomes is similar to the one adopted by Gabriel Ryan et al. for evaluating TOGA~\cite{10.1145/3510003.3510141}.

To account for the inherent variability in the responses produced by LLMs, we repeat the assertion generation for every test case 10 times. This results in situations where some assertions are correct while others are not. Hence, we adopted a threshold-based evaluation mechanism to measure the consistency and reliability of the generated test oracles across multiple runs. Specifically, we ran 10 replications of each experiment and considered three thresholds: 60\%, 80\%, and 100\%, which represent varying levels of consistency across the replications. The thresholds indicate the minimum percentage of replications that must generate the same assertion for a given test case to be considered successful. For example, the 100\% threshold is the most conservative, meaning that the test case is only considered valid if all 10 replications generate the correct  assertions, whereas the 60\% threshold is more lenient, requiring correctness in only 6 out of the 10 runs.

Additionally, during each run, if the tool fails to generate or extract an assertion from the LLM, it retries up to three times before marking the attempt as unsuccessful. This retry mechanism ensures that minor failures or inconsistencies in the LLM's response do not disproportionately affect the evaluation outcomes.

\textbf{Test Generation.} For test generation during the dataset preparation process described above, we used \evo version 1.2, to generate tests for both the original and mutated classes. We generated these tests from the buggy class and tests that passed on the buggy class but failed on the original class were selected as candidate test cases for assertion generation. The LLM used for assertion inference was prompted with context data formatted as JSON, and assertions were replaced in the test cases accordingly.

\textbf{Execution Environment.} We present here some details regarding the software and hardware used in our experiment to facilitate future replication studies.  
    The evaluation was conducted on two platforms: the TOGA (\togaun, \togawt) and API based experiments (\gptsp, \gptsp, \raggen) are executed on a machine with an Intel Core i7 processor, 32GB RAM, and Ubuntu 22.04 OS. The experiments using quantized local LLM installations (\lamasp, \lamaep, \hermessp, \hermesep) are executed on nodes in a cluster environment with AMD EPYC 7413 24-Core Processor and 0.5TB RAM. 

% latex table generated in R 4.3.3 by xtable 1.8-4 package
% Thu Oct  3 09:59:45 2024
\begin{table*}[htb]
\centering
\begingroup\fontsize{8pt}{8pt}\selectfont
\vspace{3mm}
\begin{tabular}{clrrrrr}
  \hline \\[-1.5ex]
Acceptance threshold & Variant & TP \% & FP \% & TN \% & FN \% & Failures \% \\ 
  \hline
  \hline \\[-1.5ex]
\multirow{9}{*}{60\%} & \gptep & 45.5 & 26.4 & 19.1 & 1.8 & 7.2 \\ 
   & \gptsp & 37.3 & 32.7 & 15.5 & 2.7 & 11.8 \\ 
   & \raggen & 28.2 & 46.4 & 17.3 & 4.5 & 3.6 \\ 
   & \lamasp & 10.0 & 35.5 & 13.6 & 0.9 & 40.0 \\ 
   & \lamaep & 10.9 & 25.5 & 15.5 & 0.0 & 48.1 \\ 
   & \hermessp & 9.1 & 20.0 & 8.2 & 2.7 & 60.0 \\ 
   & \hermesep & 6.4 & 22.7 & 2.7 & 1.8 & 66.4 \\ 
   & \togawt & 5.5 & 4.5 & 9.1 & 0.0 & 80.9 \\ 
   & \togaun & 8.2 & 13.6 & 25.5 & 0.0 & 52.7 \\ 
  \hline \\[-1.5ex]
\multirow{9}{*}{80\%} &\gptep & 36.4 & 20.0 & 16.4 & 0.9 & 26.3 \\ 
   & \gptsp & 34.5 & 30.9 & 13.6 & 1.8 & 19.2 \\ 
   & \raggen & 25.5 & 40.9 & 13.6 & 2.7 & 17.3 \\ 
   & \lamasp & 7.3 & 30.9 & 10.9 & 0.9 & 50.0 \\ 
   & \lamaep & 7.3 & 17.3 & 10.9 & 0.0 & 64.5 \\ 
   & \hermessp & 6.4 & 8.2 & 2.7 & 1.8 & 80.9 \\ 
   & \hermesep & 1.8 & 3.6 & 0.9 & 0.9 & 92.8 \\ 
   & \togawt & 5.5 & 4.5 & 9.1 & 0.0 & 80.9 \\ 
   & \togaun & 8.2 & 13.6 & 25.5 & 0.0 & 52.7 \\ 
  \hline \\[-1.5ex]
\multirow{9}{*}{100\%} & \gptep & 30.0 & 15.5 & 11.8 & 0.9 & 41.8 \\ 
   & \gptsp & 29.1 & 27.3 & 11.8 & 0.9 & 30.9 \\ 
   & \raggen & 18.2 & 21.8 & 7.3 & 0.9 & 51.8 \\ 
   & \lamasp & 6.4 & 25.5 & 8.2 & 0.9 & 59.0 \\ 
   & \lamaep & 7.3 & 14.5 & 10.9 & 0.0 & 67.3 \\ 
   & \hermessp & 3.6 & 5.5 & 0.0 & 1.8 & 89.1 \\ 
   & \hermesep & 0.9 & 0.0 & 0.0 & 0.9 & 98.2 \\ 
   & \togawt & 5.5 & 4.5 & 9.1 & 0.0 & 80.9 \\ 
   & \togaun & 8.2 & 13.6 & 25.5 & 0.0 & 52.7 \\ 
   \hline
   \hline
\end{tabular}
\endgroup
\caption{Summary of the performances of the different variants. True Positive (TP) are cases where the test passes on the original code but fails on the mutated. False Positive (FP) includes tests that fail on both original and mutated code. True Negative (TN) are those tests that pass on both original and buggy code. False Negative (FN) tests fail on the original code and pass on the mutated. The acceptance threshold indicates the fraction of replicas that have to show concordant results (Section~\ref{sec:evaluation_setup}). Failures group all the cases in which the generated tests do not compile or the replicas exhibit discordant behavior.}
\label{tab:performance_summary}
\end{table*}

\subsection{Results}
In this section, we present the results of the experiment along with the two research questions we outlined earlier. 
% RQ1
\subsubsection*{RQ1: Can {\toolname} infer correct oracles?}
To answer RQ1, we analyzed the correctness of the assertions generated by the different {\toolname} variants as well as by the baseline (TOGA). The primary metric for correctness was the ability of the newly generated assertion to pass on the original class and fail on the buggy class (true positives). This ensures that the assertion reflects the intended behavior of the method under test with respect to the test statements in the specific test case.

\begin{figure*}[t]
\includegraphics[width=\textwidth]{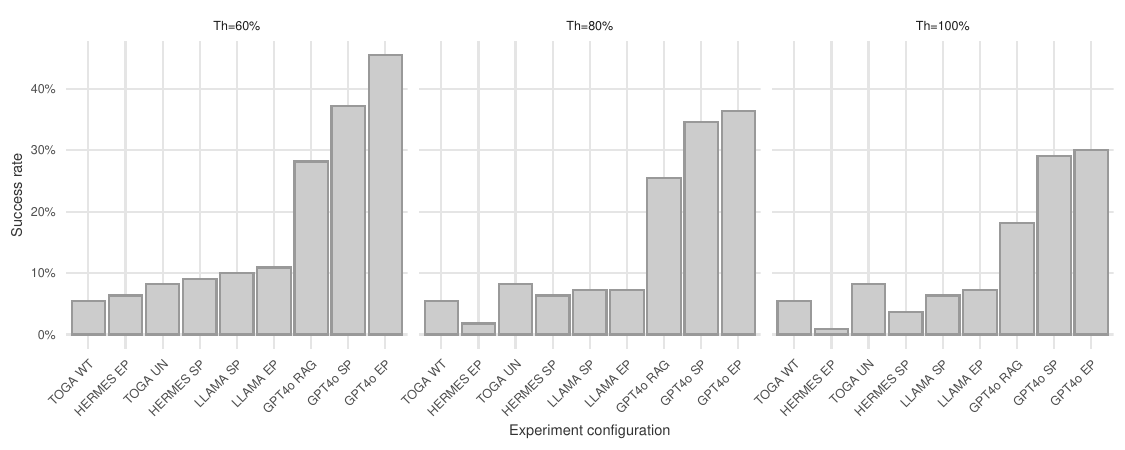}
\caption{Performance of the different variants with three values of consistency thresholds (Th).}
\label{fig:performance-of-tool}
\end{figure*}

% \begin{itemize}
    In Table~\ref{tab:performance_summary} we present the results of our experimental runs on the dataset of 203 tests. We also graphically show the success rates of the various variants in Figure~\ref{fig:performance-of-tool}. We note here that we do not report results for two of the variants described above (\ragsp and \ragep) as we excluded them from the experimentation based on the results of \raggen where the results were not as promising while the experiments are quite time consuming and expensive. Hence, in the interest of saving time and resources, we decided to exclude them from the experiment. We also further note that for similar reasons we ran the experiment with the RAG variant (\raggen) only 5 times per test case, while for all the other variants we performed 10 repetitions.  
    
    The results demonstrate that {\toolname} can indeed infer correct assertions, with the GPT-4 model performing significantly better than LLAMA and HERMES in terms of overall success rates. The use of Extended Prompts (EP) led to higher success rates than the Simple Prompt (SP), suggesting that providing more information improves assertion inference. Specifically, the \gptep variant gives the best performance out of the variants. 
    
    Considering the baseline variants (\togaun and \togawt) which employ specialized neural models, the results show that three of \toolname's variants (\gptep, \gptsp, and \raggen) perform consistently better than the baseline variants in all three levels of threshold. The baseline variants achieve low success rates as can be seen from Table~\ref{tab:performance_summary}. In particular the baseline with the default settings (\togawt) achieves low success rates, while relaxing some of the settings (as discussed earlier in Section~\ref{sec:evaluation_setup}) appears to have improved the success rate where the \togaun variant achieves 3 percentage points improvement over the default variant (\togawt) for all the threshold values considered. 
    We note here that in our dataset there were a number of test cases (as generated by \evo) that involve exception handling assertions. In these cases, none of the variants experimented with were able to infer the correct assertion. Hence, the results presented here are only for those test cases involving non-exception assertions (more on this in the Discussion Section~\ref{sec:discussion}).
% \end{itemize}

% RQ2
\subsubsection*{RQ2: Does enhancing the context improve assertion generation?}
For RQ2, we focus on the different prompt variants of {\toolname} to evaluate their relative effectiveness. As can be seen in Table~\ref{tab:performance_summary} and Figure~\ref{fig:performance-of-tool}, at 60\% threshold, {\toolname}'s Extended Prompt (EP) achieved the highest success rate of 45.5\%, compared to the Simple Prompt (SP) at 37.3\%. When we increase the threshold to 80\%, EP still outperformed other variants with 36.4\%, while SP achieved 34.5\%. In the most conservative scenario, where 100\% consistency is required across all replications, the success rates drop, but EP continues to lead at 30\%, with SP at 29.1\%. Interestingly, the RAG-based approaches consistently underperformed across all thresholds, suggesting that while enriching the LLMs with additional structured data holds potential, it requires further refinement to realize its full benefits. We also notice that the difference between the EP and SP variants diminishes as the threshold level is increased.

These results suggest that while {\toolname} can infer correct assertions, the introduction of RAG did not yield the expected improvements. In fact, the \raggen variant underperformed compared to both the Simple and Extended Prompts. The success rate for \gptep, though marginally higher than \gptsp, indicates that additional context does help but not substantially.

We also note from the results in Table~\ref{tab:performance_summary} that there are several cases of failure (Failures column) where the assertion generation did not produce candidates, mainly because the generated assertions did not compile/run successfully. This also shows an inherent issue with using LLMs, at least in their current state, which does not necessarily guarantee a valid response. Further studies are required to understand how much of this problem could be mitigated by improving the prompting mechanism, of course besides improving the performances of the models themselves. 

To facilitate replication of the work reported in this paper and further experimentation we make all data and results publicly available online~\cite{replication_package}.

% Discussion
\subsection{Discussion} \label{sec:discussion}

The experimental results demonstrate that \toolname effectively infers correct test oracles across various scenarios. As shown in Table~\ref{tab:performance_summary}, the \texttt{True Positive (TP)} column highlights the percentage of successfully inferred assertion oracles, where \toolname was able to correctly identify assertions in 110 test cases. Notably, however, none of the variants
%of \toolname 
was successful in inferring any \texttt{Exception Oracles}, which is an unexpected outcome. While TOGA is typically capable of classifying both assertion and exception oracles, its failure to infer exceptions in this case is surprising, especially given its training on Exception Oracles. This indicates potential limitations in TOGA's ability to generalize beyond its training data in practical scenarios. This finding also indicates that \toolname approach requires further refinement in future work for handling oracles involving exceptions. 

The evaluation reveals a notable performance disparity among various models in inferring correct test oracles. The \texttt{GPT-4o} model from OpenAI API consistently outperforms all other variants, demonstrating superior effectiveness in this task. However, its reliance on cloud-based processing may raise significant privacy concerns, as sensitive data may be exposed during model interaction. On the other hand, quantized models like Hermes and Llama, while not achieving comparable performance levels, provide the advantage of being run entirely locally. This eliminates privacy risks, making them suitable for environments where data confidentiality is paramount. 

One of the key finding is that providing more detailed and structured information via RAG did not lead to better oracle inference. The \texttt{RAG variants} underperformed compared to the simpler prompts, contradicting our expectation that more context would result in more accurate assertions. This could suggest that the LLM has limitations in effectively integrating complex structured data in this setting. It also suggests that, while LLMs like \texttt{GPT-4o} benefit from richer natural language context (as seen in the improvement from SP to EP), they may not gain from structured data unless integrated in a more intuitive or specialized way. While RAG shows reasonable performance, it comes at a higher operational cost compared to other variants. This highlights the trade-off users must consider: choosing between superior model performance and privacy assurance or managing costs while potentially sacrificing some effectiveness.

An interesting observation from our results is that the model occasionally generates natural language descriptions of assertions instead of source code—though unintended, this highlights the potential of LLMs for interpreting and explaining code, warranting further exploration of such capabilities.

These findings underscore the need for developers and organizations to carefully evaluate their priorities, whether they lean towards performance, privacy, or cost-efficiency, in selecting the most appropriate model for their software testing needs. The flexibility of \toolname is aimed at supporting developers in such situations.

% Threats to Validity
\subsection{Threats to Validity}
%\noindent\textbf{Internal Validity}

\textbf{Data Leakage.} A potential threat to internal validity is the possibility of data leakage, particularly with respect to the LLMs used in our study. LLMs may have been pre-trained on similar datasets, potentially leading to over-optimistic results when generating test oracles. However, to mitigate this risk, we carefully selected Java projects based on a different benchmark and generated unique mutants that are unlikely to have been seen by the model during pre-training. While this reduces the chances of data leakage, it is impossible to entirely rule out the influence of pre-training data on the results.

\textbf{Mutant Generation Bias.} Our evaluation relies on automatically generated mutants to benchmark the correctness of inferred assertions. These mutants, while useful for simulating bugs, may not fully represent real-world software faults. This introduces a potential bias in evaluating {\toolname}’s effectiveness in real-world scenarios. Future work could involve using a wider variety of bug benchmarks or real-world defects to strengthen the internal validity of the approach.

%\noindent\textbf{External Validity}

\textbf{Generalization.} The external validity of our study is primarily limited by the scope of the dataset, which consists exclusively of Java projects. While our results demonstrate {\toolname}’s ability to infer correct test oracles in this context, it is unclear how well the approach would generalize to other programming languages or frameworks. Future research should aim to evaluate {\toolname} on projects from additional programming ecosystems, such as Python, C\#, or JavaScript, to assess the generalizability of the approach.

\textbf{LLM Dependency.} Another factor affecting generalizability is the reliance on specific large language models, such as GPT-4, used for assertion inference. The performance of {\toolname} may vary across different LLMs, particularly those with different architectures or training data. As new models emerge, it will be important to assess whether the observed performance improvements hold across a wider range of LLMs and domains.

%\noindent\textbf{Construct Validity}

\textbf{Evaluation Metrics.} The success of our approach is primarily measured by the accuracy of inferred assertions in passing on the original class and failing on the mutant. While this is a strong indicator of correctness, it does not fully capture other aspects of a test's effectiveness, such as the ability to detect subtle, real-world bugs or maintainability of the generated tests. Future work could focus on broadening the evaluation metrics.

% conclusion & future work
\section{Conclusion and Future work}  \label{sec:conclusion}
In this paper we presented {\toolname}, an approach for generating test oracles using large language models. Our method demonstrates the effectiveness of leveraging context-rich prompts to guide the LLM in generating assertions for automated tests. We evaluated our approach across four variants: Simple Prompt, Extended Prompt, RAG with Generic Prompt, and RAG with Simple Prompt. The results indicate that the \emph{Extended Prompt} (EP) variant outperformed the \emph{Simple Prompt} (SP) variant, achieving an assertion correctness rate of 30\%, compared to 29.1\% for SP considering the most conservative scenario. However, the use of RAG (both RAG and RAG SP) did not lead to the anticipated performance improvements, with RAG yielding a lower correctness rate of 18.2\%. This suggests that while adding more contextual information improves the quality of generated assertions, the retrieval-based augmentation in RAG may require further optimization to fully realize its potential. The results demonstrate the potential of LLM-based test oracle generation in augmenting software testing practices, reducing manual effort in writing assertions, and improving the quality of automatically generated test cases.

Despite these promising results, several areas for future improvement remain. First, our method relies on syntactic correctness and successful compilation to validate assertions but lacks a formal assurance mechanism to verify the semantic correctness of the generated oracles. Approaches like \emph{Assured LLM-Based Software Engineering} have explored assurance layers to rigorously filter and validate LLM outputs before they are accepted~\cite{10.1145/3643661.3643953}. Incorporating a similar assurance layer could improve the reliability of generated assertions, ensuring they truly reflect the intended behavior of the system under test. Moreover, we currently lack an assertion ranking mechanism to prioritize more accurate or optimal assertions, which has been explored in related works~\cite{10.1145/3510003.3510141}. Developing such a mechanism could further enhance the quality of generated test cases. Moreover, our evaluation focused on Java projects, and future work should explore applying the approach to other programming languages and software domains. This would allow us to assess the generalizability and scalability of \toolname{} across diverse systems and testing scenarios. Finally, future work could investigate the role of AI-generated comments as a supplement to developer-written comments, assessing how they influence the quality of generated assertions and the overall performance of the approach. By analyzing the synergy or redundancy between these comment types, we could gain insights into optimizing prompt engineering and improving the contextual understanding of large language models in software testing.

\section*{Acknowledgment}
We acknowledge the support  of the PNRR project FAIR -  Future AI Research (PE00000013),  under the NRRP MUR program funded by the NextGenerationEU.

% \printbibliography
\bibliographystyle{IEEEtran}
\bibliography{references}

\end{document}